\documentstyle[12pt,aps,prl]{revtex}

\begin {document}

\title {Fermi Liquid Properties of a Two Dimensional Electron System
        With the Fermi Level Near a van Hove Singularity}

\author {D. Menashe  and B.
         Laikhtman}

\address {Racah Institute of Physics,
          Hebrew University, Jerusalem 91904, Israel}

\maketitle

\begin{abstract}

We use a diagrammatic approach to study low energy physics
of a two dimensional electron system where the Fermi level is near
van-Hove singularies in the energy spectrum. We find that in most
regions of the $\epsilon_F$-$T$ phase diagram the system
behaves as a normal Fermi liquid rather than a marginal Fermi liquid.
Particularly, the imaginary
part of the self energy is much smaller than the excitation energy,
which implies well defined quasiparticle excitations, and single
particle properties are only weakly affected by the presence of
the van-Hove singularities. The relevance
to high temperature superconductivity is also discussed.

\end{abstract}


Since the discovery of high temperature superconductors (HTSC's), the
role played by a saddle point in the energy spectrum, known as a
van-Hove singularity (vHS), has been actively debated in the
literature \cite
{Hirsch 86,Dzyalo 87,Tsuei 90,Lee 87,Gop 92,Gon 96,Dzyalo 96,Lu 96}.
The experimental observation of vHS's near 
the Fermi surface in many HTSC's \cite{Lu 96} has led to 
the so called van-Hove scenario to explain the normal state properties of
these materials \cite {Tsuei 90,Lee 87,Gop 92,Gon 96,Dzyalo 96},
as well as the
superconducting properties \cite {Hirsch 86,Dzyalo 87,Tsuei 90}.
In this scenario, the presence of the vHS's, as well as
weak electron electron (e-e) interactions, are used to
explain the main body of experimental data.

One of the more notable properties of the normal state is
the linear temperature dependence of the resistivity  \cite{Lu 96}.
This has been explained within 
the van-Hove scenario by arguing that when the temperature is
larger than the energy difference between the Fermi level and the
vHS's, the e-e scattering rate is linear in temperature
\cite{Tsuei 90,Lee 87,Gop 92}.
More generally, in this temperature region the imaginary part of
the self energy has
been claimed to have the behavior
${\rm Im}\Sigma\left({\bf k},\epsilon\right) \propto
\max(\epsilon,T)$ for
a large range of energies $\epsilon$,
thus explaining  many
other normal state properties \cite{Varma 89}. From a
theoretical view, the linear energy dependence of
${\rm Im}\Sigma\left({\bf k},\epsilon\right)$ differs from
the quadratic dependence of a regular Fermi liquid (FL), prompting
some workers to classify the system as a marginal Fermi
liquid (MFL) \cite{Varma 89}.

Recently, it has been suggested \cite{Dzyalo 96} that the situation
is more complicated, and that in a certain region of the
$\epsilon_F$-$T$ phase diagram
($\epsilon_F$ is relative to the vHS's),
FL theory breaks down altogether and a non Fermi liquid
(NFL) exists instead. The resulting
phase diagram is shown in Fig. \ref {Phase_Diagram}, and
can be summarized as follows: When $\epsilon_F$ is small 
enough, or $T$ is low enough
(specific criteria will be given further on),
the system behaves as a NFL. In other regions
of the phase diagram the system behaves as either a regular
FL or a MFL:
When $T < |\epsilon_F|$, the affect of the vHS's is not felt and
${\rm Im}\Sigma \propto T^2$,
indicating a regular FL behavior. In the
opposite case the temperature is high enough so that
quasiparticles near the Fermi level feel the affect of the vHS's,
resulting in the linear temperature dependence of
${\rm Im}\Sigma$ characteristic of a MFL.

In the present work we use a standard diagrammatic approach to
study the low energy physics of a system of weakly interacting
electrons with the Fermi level near vHS's.
We show that the
leading order perturbation calculation of the self energy
\cite{Tsuei 90,Lee 87,Gop 92} is invalid, and instead we need to
renormalize the self energy by summing diagrams in the particle-particle
channel.
Dzyaloshinskii \cite{Dzyalo 96} used a more complex version of this
scheme to show the existence of a NFL in the
bottom left corner of the phase diagram in Fig. \ref{Phase_Diagram}.
However, we
concentrate rather on the implications for the Fermi liquid properties
in the region not occupied by a NFL.
For this region, our summation scheme is strictly valid, and  our main result
is that the energy dependence of
${\rm Im}\Sigma\left({\bf k},\epsilon\right)$ is {\sl always} weaker
than linear, even in the part of the phase diagram previously
thought to contain a MFL. Furthermore, we show that the
quasiparticle properties, namely the effective mass and
the quasiparticle weight, are only weakly affected
by the presence of the vHS's.

These results imply that the van-Hove scenario in it's present
form cannot explain fully the normal state properties of HTSC's.
These include the linear temperature dependence of the
conductivity \cite{Lu 96}, the suppression of the
quasiparticle weight \cite{Varma 89},
and the extended nature of most vHS's \cite{Lu 96}.
Theoretically, the results are important since 
${\rm Im}\Sigma\left({\bf k},\epsilon\right) \ll \epsilon $,
so that quasiparticles are well defined
excitations, as required by FL theory. Thus, we conclude that
regular FL theory is valid in {\sl all} parts of the phase 
diagram not occupied by the NFL.

The basic system we discuss here is that of a square lattice with an
arbitrary energy spectrum. Due to symmetry,
any band of the energy spectrum will have two equal energy
saddle points per Brillouin zone (located 
at wave vectors $(0,\pi/a)$ and $(\pi/a,0)$, $a$
being the lattice constant).
When these points are close to the Fermi level,
we shall see that the low energy
physics of the system is dominated by electrons occupying regions
in ${\bf k}$-space in their vicinity.
This enables us to describe the system using the
model Hamiltonian
\begin {eqnarray} \label {Model}
     H  & = &
     \sum_{i,{\bf k},\sigma}\epsilon^{(i)}_{{\bf k}}
     c^{(i)\dagger}_{{\bf k},\sigma}c^{(i)}_{{\bf k},\sigma} +
     \frac{g}{2S}
     \sum_{i,j,{\bf k},{\bf k}',{\bf q},\sigma,\sigma '}
     c^{(i)\dagger}_{{\bf k}+{\bf q},\sigma}
     c^{(j)\dagger}_{{\bf k'}-{\bf q},\sigma '}
     c^{(j)}_{{\bf k'},\sigma '}  c^{(i)}_{{\bf k},\sigma} +
     \nonumber \\ & &
     \frac{\tilde{g}}{2S}
     \sum_{i \neq l,t \neq j,{\bf k},{\bf k}',{\bf q},\sigma,\sigma'}
     c^{(i)\dagger}_{{\bf k}+{\bf q},\sigma}
     c^{(t)\dagger}_{{\bf k'}-{\bf q},\sigma '}
     c^{(j)}_{{\bf k'},\sigma '}  c^{(l)}_{{\bf k},\sigma}.
\end {eqnarray}
Here $S$ is the sample area, and 
$c^{(i)\dagger}_{{\bf k},\sigma}$  $(c^{(i)}_{{\bf k},\sigma})$
creates (annihilates) electrons with spin $\sigma$ and wave vector
${\bf k}$ relative to saddle point $i (=1,2)$. The spectrum
near these points is
\begin{eqnarray} \label{Spectrum}
     \epsilon^{(1)}_{{\bf k}} & = &
     t_{x} k_{x}^{2} - t_{y} k_{y}^{2},
     \quad \quad 
     \epsilon^{(2)}_{{\bf k}} = 
     t_{y} k_{x}^{2} - t_{x} k_{y}^{2},
\end{eqnarray}
where $t_{x} \equiv \hbar^{2}/2m_x$ , $t_{y} \equiv \hbar^{2}/2m_y$
and $m_{x},m_{y}$ are the effective masses.
If $\left|t_{x}-t_{y}\right| \ll t_{x},t_{y}$,
there is said to be significant nesting between the two saddle
point. This is not the general case however, so 
throughout this work we consider
only the non-nested case, meaning
$\left|t_{x} - t_{y}\right| \sim t_{x} \sim t_{y}$.
$g$ ($\tilde {g}$) is a  ${\bf k}$-independent coupling
constant representing  processes in which the
initial and final saddle points of a given electron are the
same (different).
Strictly, each of these constants should be split into two,
depending on whether or not both incoming electrons are at the
same saddle point. However, we will show that only processes
where both incoming electrons are at the same saddle point are
relevant, thus allowing us to ignore this point.
To complete the definition of our model we assume a wave vector
cutoff $k_c \sim 1/a$, which defines
an energy cutoff
$\epsilon_{c} \sim \max\left(t_{x},t_{y}\right) k^{2}_{c}$.

To understand the basic physics of this model we first
consider a simpler model with only one saddle point
instead of two. Thus, we omit  the
$\tilde{g}$ coupling term
in the Hamiltonian, and have only one
species of electrons whose spectrum is given by 
$\epsilon_{\bf k}^{(1)}$ in Eq. (\ref {Spectrum}). 
For this model it is
convenient to perform a change of variables to
$\epsilon_{{\bf k}} = t k_{x} k_{y}$,
where $t \equiv \sqrt{t_{x}t_{y}}$.
Our original assumption regarding $t_{x}$ and $t_{y}$ means that
$t \sim t_{x},t_{y}$ and that the new form of the spectrum also has a
cutoff $\sim k_{c}$.
The density of states (DOS) for this spectrum
is $D\left(\epsilon\right) \approx \left(1/2\pi^{2}t\right)
\ln\left(t k_{c}^{2}/\epsilon\right) \approx
\left(1/2\pi^{2}t\right) \ln\left(\epsilon_{c}/\epsilon\right)$, which
diverges at small $\epsilon$, leading to the so called vHS.
This means that the main contribution to the DOS
comes from states lying close to the saddle point,
which justifies our original assumption that
the low energy physics is dominated by such states.
Furthermore, a simple estimate shows that an electron with energy
$\epsilon \ll \epsilon_{c}$ will typically have a wave vector
${\bf k}$ satisfying
\begin {eqnarray} \label {Logs}
     \ln\frac{k_{c}}{\left|k_{x}\right|} \sim
     \ln\frac{k_{c}}{\left|k_{y}\right|} \sim
     \ln\frac{\epsilon_{c}}{\left|\epsilon\right|} \gg 1.
\end {eqnarray}
The above equation, which means that all logarithmic factors appearing
in our model are large and of the same order,
is central to this work and will be used throughout.

We  begin our treatment of the simpler model
by considering leading order contributions to
$\Sigma\left({\bf k},\epsilon\right)$,
which come from the diagrams in Fig.
\ref{Diagrams}a and \ref{Diagrams}b.
When $|\epsilon| < |\epsilon_F|$, it is easy to show
that these diagrams give regular FL behavior,
meaning ${\rm Im} \Sigma \propto \epsilon^2$.
Instead, we concentrate
on the more interesting case
$|\epsilon|  > |\epsilon_F|$. Also, for 
simplicity we neglect finite temperature effects, which 
serve only to replace $\epsilon$ with $\sim T$ when $|\epsilon| < T$.
Under these conditions, we obtain
\begin{eqnarray} \label{ImSigma.E1}
   {\rm Im}\Sigma\left({\bf k},\epsilon\right) & = &
   \left\{
   \begin{array}{ll}
      \displaystyle
      \frac{-g^2\epsilon}{4\pi^{3} t^{2}}  &
      \displaystyle
      \quad {\rm for} \quad \epsilon = \epsilon_{\bf k} \\
      & \\
      \displaystyle
      \frac{-g^2\epsilon}{8\pi^{3} t^{2}}
      \left( \ln 2 + (1- \ln 2 )\frac{\epsilon_{\bf k}}{\epsilon}
      \right)
      \ln\left(\frac{\epsilon_c}{\epsilon}\right) &
      \displaystyle
      \quad {\rm for} \quad |\epsilon| \gg |\epsilon_{\bf k}|
   \end{array}
   \right.
\end{eqnarray}
The main point here is that the on shell self energy is linear in
the energy, which is a signature of MFL behavior.
Using the analytic properties of the self energy,
we may calculate it's real part. Since this is
logarithmically large, it's main contribution is determined
by ${\rm Im}\Sigma\left({\bf k},\epsilon\right)$ with
$|\epsilon| \gg |\epsilon_{\bf k}|$, so that
\begin{eqnarray} \label{ReSigma.E1}
     {\rm Re}\Sigma\left({\bf k},\epsilon\right) & = &
     \frac{-g^{2}} {8\pi^{4} t^{2}}
     \left[ (\ln 2) \epsilon + (1 - \ln 2) \epsilon_{\bf k}\right]
     \ln^{2}\left(\frac{\epsilon_c}{\max(|\epsilon_F|,|\epsilon|)}
            \right).
\end{eqnarray}
We note that this result is equivalent to that obtained
by Dzyaloshinskii \cite {Dzyalo 96}, who calculated
${\rm Re}\Sigma\left({\bf k},\epsilon\right)$ directly.
Also, it is a further
indication of MFL behavior, since it means
that the quasiparticle properties are strongly renormalized in the
low energy limit.

We next consider higher order contributions to the self energy.
It has been shown \cite{Abrikosov 63} that all such contributions
may be accounted for by correcting the bare
Greens functions of diagrams \ref{Diagrams}a and
\ref{Diagrams}b, or by replacing one of the interaction
lines with
a vertex. We will show shortly that corrections to the bare Greens
functions are small, thus leaving us with the vertex replacements.
The possible vertices are shown in  Fig.
\ref{Diagrams}c-\ref{Diagrams}e, and an example of
a third order diagram including such
vertices is shown in diagram \ref{Diagrams}f.
The singular nature of these vertices has been discussed
many times within the
context of the van-Hove scenario
\cite{Hirsch 86,Dzyalo 87,Gon 96}.
It has been shown that the particle-particle channel has a square
logarithmic divergence, i.e.
$ \Gamma_{c} \sim t^{-1} g^{2}\ln^{2}(\epsilon_c/\epsilon)$,
whereas the particle-hole channel only diverges as a single logarithm,
i.e. $ \Gamma_{d} \sim \Gamma_{e} \sim t^{-1} g^{2}
\ln(\epsilon_c/\epsilon)$
(here  $\epsilon$ is the scale of the incoming energies of the vertices,
and the subscripts $c$, $d$ and $e$
refer to Fig. \ref{Diagrams}).
Thus, at low enough energies, we may neglect the particle-hole
channel, and need only consider
corrections to the self energy coming from the particle-particle
channel, $\Gamma_c$.
Furthermore, for energies such that
$(g/t)\ln^{2}(\epsilon_c/\epsilon) > 1$, these corrections are
larger then the leading order contribution to the self energy, thus
signaling the breakdown of leading order perturbation theory.

The above arguments can easily be extended to higher order diagrams,
which means that the self energy must be renormalized by summing the
particle-particle ladder \cite{Menashe 96} 
 and inserting it into diagrams
\ref{Diagrams}a and \ref{Diagrams}b, as shown 
in Fig. \ref{SingleLadder}.
For $|\epsilon| < |\epsilon_F|$ we still obtain
${\rm Im}\Sigma\left({\bf k},\epsilon\right)\propto \epsilon^2$,
however, the important effect of the renormalization can
be seen in the opposite case, when $|\epsilon| > |\epsilon_F|$.
Here we obtain
\begin{eqnarray} \label{ImSigma.E2}
   {\rm Im}\Sigma\left({\bf k},\epsilon\right) & = &
   \left\{
   \begin{array}{ll}
      \displaystyle
      \frac{-g^2\epsilon}{4\pi^{3} t^{2}}
      \left[\frac{1}{1+(g/8\pi^2 t)\ln^2(\epsilon_c/\epsilon)}
      \right]^2  &
      \displaystyle
      \quad {\rm for} \quad \epsilon = \epsilon_{\bf k} \\
      & \\
     \displaystyle
      \frac{-g\epsilon}{4\pi t}
      \left( \ln 2 + (1- \ln 2 )\frac{\epsilon_{\bf k}}{\epsilon}
      \right)\frac{1}
      {\ln\left(\epsilon_c/\epsilon\right)} &
      \displaystyle
      \quad {\rm for} \quad |\epsilon| \gg |\epsilon_{\bf k}|
   \end{array}
   \right.
\end{eqnarray}
which is the main result of our work.
From it, it is clear that as long as
$(g/t)\ln^2(\epsilon_c/\epsilon) > 1$, the on shell self energy
satisfies
${\rm Im}\Sigma\left({\bf k},\epsilon\right) \sim
\epsilon / \ln^{4}(\epsilon_c/\epsilon) \ll \epsilon$. Thus
even when $|\epsilon| > |\epsilon_F|$ Landau quasiparticles are
well defined, and the system behaves as a regular FL rather than
a MFL. Using analyticity, we obtain for the  real part 
of the self energy
\begin{eqnarray} \label{ReSigma.E2}
     {\rm Re}\Sigma\left({\bf k},\epsilon\right) & = &
     \frac{-g} {2\pi^{2} t}
     \left[ (\ln 2) \epsilon + (1 - \ln 2) \epsilon_{\bf k}\right]
     \ln\left(\ln\left(\frac{\epsilon_c}{\max(|\epsilon_F|,|\epsilon|)}
                \right)
        \right).
\end{eqnarray}
Comparing to Eq. (\ref{ReSigma.E1}), we see that the
real part of the self energy is strongly renormalized, so the
affect of the vHS is much weaker then in the leading order case.
Using Eq. (\ref{ReSigma.E2}) to calculate corrections
to the quasiparticle weight and effective mass, we see they
are $\sim (g/t)\ln(\ln(\epsilon_c/\epsilon))$.
Since our renormalization procedure is valid for the
region $(g/t)\ln^2(\epsilon_c/\epsilon) > 1$, these corrections can
generally be considered small. Furthermore,
in most samples $|\epsilon_F|$ is finite (see below), 
so the correction are
small even for very low energies.
This justifies our not including 
corrections to the bare Green's function
in our renormalization scheme.

However, it is clear that the energy $\xi$, defined approximately
by $(g/t)\ln(\ln(\epsilon_c/\xi)) \sim 1$, is an energy scale below
which our approximation becomes invalid.
As long as $|\epsilon_F| > \xi$, we may apply our results all the
way down to $T = 0$.
When $|\epsilon_F| < \xi $, we may assume our results describe the
system as long as $T > \xi$.
This estimate for the breakdown of our approximation is
consistent with the work of Dzyaloshinskii \cite{Dzyalo 96}.
There, by renormalizing the
Greens function in a self consistent manner, it
was shown that the quasiparticle weight becomes identically zero
at an energy defined by
$(g\ln2/t\pi^2)\ln(\ln(\epsilon_c/\xi)) = 1$. This signifies the
onset of the NFL in 
the bottom left hand corner of
Fig. \ref{Phase_Diagram}.
Thus, we conclude that our results 
remain valid as long as FL theory remains stable, i.e. as long as
$\max\left(T,\left|\epsilon_F\right|\right) > \xi$.

At this point it is instructive to consider typical values of
$\epsilon_c/\epsilon$ for the Cuprates.
Since the energy distance of the vHS's from the Fermi level is
typically $10-20$ mev \cite{Lu 96},
and $\epsilon_c \sim 1$ ev,
we obtain $\epsilon_c/|\epsilon_F| \sim 50-100$, so
the factor $\ln^4(\epsilon_c/\epsilon)$ appearing in the
denominator of ${\rm Im}\Sigma$ can be as large as $\sim 100$.
Furthermore, due to the high power of the logarithm, the dependence of
${\rm Im}\Sigma$ on $\epsilon$ should easily be distinguishable from
a linear dependence. Thus, the linear temperature dependence
of the resistivity cannot simply be explained by the presence of the vHS's.
On the other hand, for the factor
$(g/t)\ln(\ln(\epsilon_c/\epsilon))$ appearing in
${\rm Re}\Sigma$ to be large, we would need say, 
$\ln(\ln(\epsilon_c/\epsilon)) > 2$ (assuming weak to intermediate
coupling). This in turn means that $\epsilon_c/|\epsilon| > 10^3$,
or $|\epsilon| < 1$ mev. Even in samples with very small
$|\epsilon_F|$, such an energy scale would probably not
be observable due to c-axis coupling, or other smearing 
mechanisms. Therefore,
the presence of the vHS is unlikely to explain the
strong renormalization of quasiparticle properties observed in the
Cuprates.

Next we consider the full model given by Eq. (\ref{Model}).
Besides the intra saddle point
processes already discussed, we now include coupling 
between the saddle points (inter saddle point processes).
An example of a leading order inter saddle
point contribution to the self energy
is shown in Fig. \ref{Diagrams}g. For the non-nested case we
consider here, it can be shown that all such contributions are
smaller than the intra saddle point ones discussed
earlier \cite{Tsuei 90}.
What of higher order inter saddle point contributions, 
which can be constructed by inserting inter saddle point
vertices into diagrams \ref{Diagrams}a and
\ref{Diagrams}b?
Of the various inter saddle point vertices, only the
particle-particle vertex shown in Fig. \ref{Diagrams}h has the necessary
$\ln^2(\epsilon_c/\epsilon)$ energy dependence to compete with 
vertex \ref{Diagrams}c.
Thus, we construct the modified ladder approximation 
shown in Fig. \ref{DoubleLadder}, and
obtain for the renormalized scattering vertex 
\begin{eqnarray} \label {DoubleGamma}
     \Gamma
     & \approx &
     \frac{g + g^{2} a \Lambda^{2} +  \tilde{g}^{2} a \Lambda^{2}}
          {1+ g a\Lambda^{2} + \tilde{g} a \Lambda^{2} +
              \left(g a \Lambda^{2} \right)^{2} -
              \left(\tilde{g} a \Lambda^{2} \right)^{2}  },
\end{eqnarray}
where $\Lambda \equiv \ln(\epsilon_c/\epsilon)$, and
$a \equiv 4/\left(\left(2\pi\right)^{2}t\right)$.
We see that as long as $g>\tilde{g}$ we still have
$\Gamma \propto \Lambda^{-2}$ for large $\Lambda$, and 
all our previous results for the self energy remain
valid to within prefactors of the logarithmic factors (which
now depend on $\tilde{g}$ as well as $g$).
If $g < \tilde{g}$ then the denominator of
Eq. (\ref{DoubleGamma}) becomes zero for large enough $\Lambda$, leading
to a BCS type instability first pointed out by
Gonzalez {\sl et al} \cite{Gon 96}.
Here we do not dwell on the question of
whether or not $g<\tilde{g}$, we merely note that
$\tilde{g}$ represents processes with large
momentum transfer, whereas $g$ represents
processes with small momentum transfer. Therefore it is
reasonable to assume that $g > \tilde{g} $ and that no
instability develops.

In conclusion, we have shown that in all regions of the phase
diagram where Fermi liquid theory remains valid, the system
behave as a regular FL, rather than a MFL.
This is based on the fact that
${\rm Im} \Sigma\left({\bf k},\epsilon \right) \ll \epsilon$
even in the region previously thought to be occupied by
a MFL. Furthermore, the effect of the vHS's on the quasiparticle
properties, reflected by the real part of the self energy,
has been shown to be weak.
Our work implies that the van-Hove scenario, in it's present form,
cannot explain all the normal state properties of high temperature
superconductors.

These results differ not only from the leading order 
perturbation results \cite{Tsuei 90,Lee 87,Gop 92},
but also from some more recent attempts to go beyond
leading order perturbation theory \cite{Gon 96,Beere 98}.
The renormalization procedure used 
is strictly valid as long as
FL theory remains valid, meaning
we are not near
the non Fermi liquid region of the phase diagram in
Fig. \ref{Phase_Diagram}.
For this region, the more complex theory due
to Dzyaloshinskii \cite{Dzyalo 96} should be used.

We would like to thank V. Zevin for useful discussions and for
reading this manuscript. This research was supported by The S.A.
Schonbrunn Research Exndowment Fund.

\newpage

\begin{figure}

\caption{A schematic phase diagram,
as reflected by previous works. Here we show that in
the region labeled as a MFL, the system actually behaves as a
regular FL.
$\epsilon_F = 0$ means that the Fermi level 
coincides with the energy of the vHS's.}

\label{Phase_Diagram}

\end{figure}

%%%%%%%%%%%%%%%%%%%%%%%%%%%%%%%%%%%%%%%%%%%%%%%%%%%%%%%%%%%%%%%%%%

\begin{figure}

\caption{Diagrams used in this work.
Solid (double) lines represent
electron propagators near the first (second) saddle point, whereas
dashed (double) lines represent $g$
($\tilde{g}$) interaction lines. Diagrams (a)-(f) refer to 
both the full model and the single saddle point model, whereas
diagrams (g) and (h) refer only to the full model.
(a) and (b) are leading
order diagrams for the self energy;
(c), (d) and (e) are vertices used to contruct higher order 
self energy diagrams;
(f) is an example of such a third order diagram;
(g) is an example of a leading order inter saddle point 
self energy diagram; and (h) is the only inter saddle point vertex
which needs to be included in the renormalization of the self energy.}

\label{Diagrams}

\end{figure}

%%%%%%%%%%%%%%%%%%%%%%%%%%%%%%%%%%%%%%%%%%%%%%%%%%%%%%%%%%%%%%%%%%

\begin{figure}

\caption{The ladder approximation for the self energy in the single
saddle point model. $\Gamma$ is the renormalized scattering vertex.}

\label{SingleLadder}

\end{figure}

%%%%%%%%%%%%%%%%%%%%%%%%%%%%%%%%%%%%%%%%%%%%%%%%%%%%%%%%%%%%%%%%%%

\begin{figure}

\caption{The ladder approximation for the full
model. $\Gamma$ is the
renormalized vertex corresponding to $g$,
whereas $\tilde{\Gamma}$ corresponds to $\tilde {g}$.}

\label{DoubleLadder}

\end{figure}

\end {document}